# Fabry-Perot Resonance of Bilayer Metasurfaces

G. Alagappan[*,1], F. J. García-Vidal[2], and C. E. Png[1]

[1]Institute of High-Performance Computing, Agency for Science, Technology, and Research (A-STAR), Fusionopolis, 1 Fusionopolis Way, #16-16 Connexis, Singapore 138632

[2]Departamento de Física Teórica de la Materia Condensada and Condensed Matter Physics Center (IFIMAC), Universidad Autónoma de Madrid, E-28049 Cantoblanco, Madrid, Spain

*Corresponding author: gandhi@ihpc.a-star.edu.sg

**ABSTRACT**: In this study, we constructed a Fabry-Perot cavity with nanostructured, thin resonant metasurfaces as meta-mirrors. We developed a temporal coupled-mode theory and provided an accurate generalization of Fabry-Perot resonance and analytically derived the transmission characteristics. The presence of metasurface mirrors introduces a substantial group delay, causing the field concentration to shift from the center of the Fabry-Perot cavity toward the metasurface region. This shift is accompanied by a significant increase in the quality factor of the FP resonance. In the frequency space, there are singular points where the quality factor increases exponentially. These singular points in meta-mirror cavities exist even when the cavity separations are smaller than the cavity length of the fundamental mode in the standard cavities. We discover two characteristic cavity separations, $L_c$ and $L_Q$, that differentiate the resonance in terms of lineshapes and the dominance of the quality factor. When $L < L_c$, there is strong evanescent interaction between the two metasurface mirrors, and the coupling of this interaction with the traditional resonance produces sharp Fano-shaped transmission peaks. When $L_c < L < L_Q$, we have induced transparency peaks with lorentzian line-shapes and length-independent quality factors. This length-independence enables, the meta-mirror cavity to outperform the traditional cavities by achiveing high quality factor despite a shorter cavity length.

Light travels at an incredible speed of 300,000 kilometers per second. This high speed underscores the importance of localizing and slowing down light to effectively control and manipulate light propagation and its interaction with matter. One of the key instruments that aids in this task is a Fabry-Perot (FP) cavity, which has a wide range of applications in spectroscopy [1], laser resonator design [2], optical filtering [3], frequency metrology [4], gas sensing [5], atomic [6], and astronomical optics [7].

Traditionally, FP cavities are designed with two parallel reflective solid surfaces, typically made of metallic or dielectric materials, which do not resonate. When the distance between these non-resonating mirror surfaces matches the resonance condition, the reflected light beams interfere coherently, leading to the formation of standing wave patterns within the cavity. This concentrates the energy at the center of the FP cavity, away from the mirror surfaces. But what happens when we replace these reflective solid surfaces with resonating metasurfaces (i.e., meta-mirrors)?

Metasurfaces are quasi two-dimensional nanoscale metamaterials that have enabled flat-optics technology [8-9]. The key ability of metasurfaces to modify the amplitude, phase, and polarization state of light through a thin planar layer has revolutionized conventional optical methods of wavefront shaping, beam steering, polarization manipulation, and resonance control. A wide range of novel classes of ultrathin planar optical devices such as metalenses [10], meta-gratings [11], meta-holograms [12], mode converters [13], and nanoprints [14] have been demonstrated. Here, we consider dielectric resonant metasurface. Dielectric metasurfaces offer an attractive platform for semiconductor photonics applications as they eliminate the intrinsic losses associated with plasmonic metasurfaces [15-17] in the optical regime, thereby increasing the efficiency of the metasurface. They can be fabricated using conventional photonic fabrication technologies and easily integrated into photonic integrated circuits.

We constructed a Fabry-Perot cavity utilizing thin dielectric resonant metasurfaces as meta-mirrors. Each metasurface displays a resonant transmission dip characterized by a coefficient of finesse $F$ and a quality factor $Q_s$. Employing a temporal coupled-mode formalism, we derived the transmission characteristics of the meta-mirror FP system analytically and provided an intuitive generalization of the resonance conditions applicable to both short and long FP cavities. There are singularity points in the frequency space at which the quality factor increases exponentially. These singularity points exists even for cavity separations smaller than the cavity length of the fundamental mode in the standard cavities. We discovered two characteristic cavity lengths that differentiate resonances in short and long cavity regimes. In the short cavity regime, significant evanescent interaction occurs between the two metasurfaces. This interaction, coupled with the resonance induced by the cavity length, results in sharp Fano-shaped transmission peaks. Conversely, in the long cavity regime, where the evanescent interaction becomes negligible, we observe induced transparency peaks exhibiting ultra-high-quality factors within the original spectrum. The quality factor of meta-mirror FP cavities has two contributions. One is length-independent, and another is length-dependent. The long cavities can thus be further categorized according to the dominance of the quality factor. Assuming a Lorentzian transmission profile and considering a cavity length of a few wavelengths, we have demonstrated that the quality factor scales as a function of $Q_s\sqrt{F}$, reflecting the characteristics of a single metasurface. This is in contrast to the conventional FP cavity, where the quality factor of the resonance peak scales with the length of the cavity. In the meta-mirror FP cavity, the electromagnetic field concentrates predominantly at the thin meta-mirrors region rather than at the middle of the cavity. Understanding these unique properties of light localization in FP metamirror cavities is thus essential for optimizing the performance of such cavities in various applications.

Resonant dielectric metasurfaces have been extensively researched and are defined by a periodic array of dielectric scatterers with a period $a$. Let's begin by examining a single layer of a resonant metasurface exhibiting a transmission dip at an angular frequency $\omega_s$. In a silicon metasurface with a typical membrane thickness of 220 nm embedded in silica, isolated transmission dips can be observed across a broad range of unit cells featuring circular pillars with radii ($r$) ranging from $0.1a$ to $0.17a$. These transmission dips are attributed to isolated magnetic resonances and exhibit Lorentzian line-shapes [18]. Similar to other resonant systems in photonics, this isolated resonance can be characterized using a temporal coupled mode formalism [19-23].This formalism establishes relationships between the electric field amplitude, $A$, within the metasurface and the amplitudes of the incoming $[s_1, s_2]$ and outgoing electric fields $[q_1, q_2]$ {see Figure 1(a)}. The rate of amplitude growth within the metasurface is described by the equation $\frac{dA}{dt} = (j\omega_s - \Gamma)a + c_s s_1 + c_s s_2$, where $\Gamma$ represents the decay rate, and $c_s$ is the coupling coefficient accounting for the coupling of incoming waves $s_1$ and $s_2$ from both sides of the metasurface. The outgoing amplitudes are given by $q_1 = e^{j\gamma}s_2 + c_a A$ and $q_2 = e^{j\gamma}s_1 +$

$c_a A$ where $\gamma$ is a phase factor. Without metasurface or resonance mechanism, we have $q_2 = s_1$, so $\gamma$ is an acquired phase when the incident light passes the metasurface. For single metasurface, the transmission is independent of $\gamma$. However, it is critical in determining the transmission characteristics of the bilayer system. Utilizing time reversal symmetry, it can be demonstrated that $c_s c_a = -e^{j\gamma}\Gamma$ and $|c_a|^2 = \Gamma$ [see supplementary information]. The electric field transmission and reflection coefficients can be derived by applying Fourier transforms to the aforementioned equations. For a single metasurface, the electric field amplitiude transmission and reflection coefficients can be expressed as $t_s(\omega) = \frac{q_2}{s_1}\big|_{s_2=0} = \frac{j\delta e^{j\gamma}}{j\delta + \Gamma}$ and $r_s(\omega) = \frac{q_1}{s_1}\big|_{s_2=0} = -\frac{\Gamma e^{j\gamma}}{j\delta + \Gamma}$, respectively. Here, $\delta = \omega - \omega_s$ where $\omega = 2\pi c/\lambda$ is the angular frequency with $c$ and $\lambda$ being the speed of light and the freespace wavelength, respectively. Note that the phase of $t_s$ switches between $-\pi$ to $\pi$ at $\omega = \omega_s$. The decay rate $\Gamma$ (due to the coupling to the external radiation modes) can be accurately estimated from the numerically evaluated transmission spectrum of the metasurface and it is given by $\Gamma = \frac{\omega_s}{2Q_s}$. In Figure 1(b), we plot the transmission spectrum of the silicon metasurface with $r = 0.15a$. For an easy scaling, the horizontal axis represents normalized frequency, $a/\lambda = \omega a/(2\pi c)$. The figure demonstrates the good agreement between the analytical expressions derived from the temporal coupled mode formalism and three-dimensional (3D) finite-difference time-domain (FDTD) calculations [24-25]. In Fig. 1(b), the resoance has a quality factor $Q_s \cong 150$.

For a bilayer of resonant metasurfaces separated by a distance $L$, we can develop the temporal coupled mode equations by assuming that the field amplitude in each metasurface as $A_1$ and $A_2$. In isolation, the electric field amplitudes of the two metasurfaces evolve independently as described by the earlier equations of a single metasurface. However, when the metasurfaces are brought together to form a parallel Fabry-Perot cavity [refer to Figure 1(c)], the electric field amplitudes couple and interfere with each other. The evolution of $A_1$ in the first metasurface can be written as: $\frac{dA_1}{dt} = (j\omega_s - \Gamma)A_1 + c_s s_1 + c_s s_2'' + j\mu A_2$. Here, $\mu$ is the coupling coefficient that accounts for the evanescent coupling of the two metasurfaces. The incoming electric field on the right side of the first metasurface, denoted as $s_2''$, can be expressed as: $s_2'' = e^{-jkL} s_2'$. Here, $kL$ with $k = \frac{n\omega}{c}$ and $n$ being the refractive index of the ambient, constitutes to the phase accumulated by the transmitted wave from the second metasurface upon reaching the first metasurface, and $s'_2$ is the transmitted amplitude in the second metasurface [see Fig. 1(c) for details].

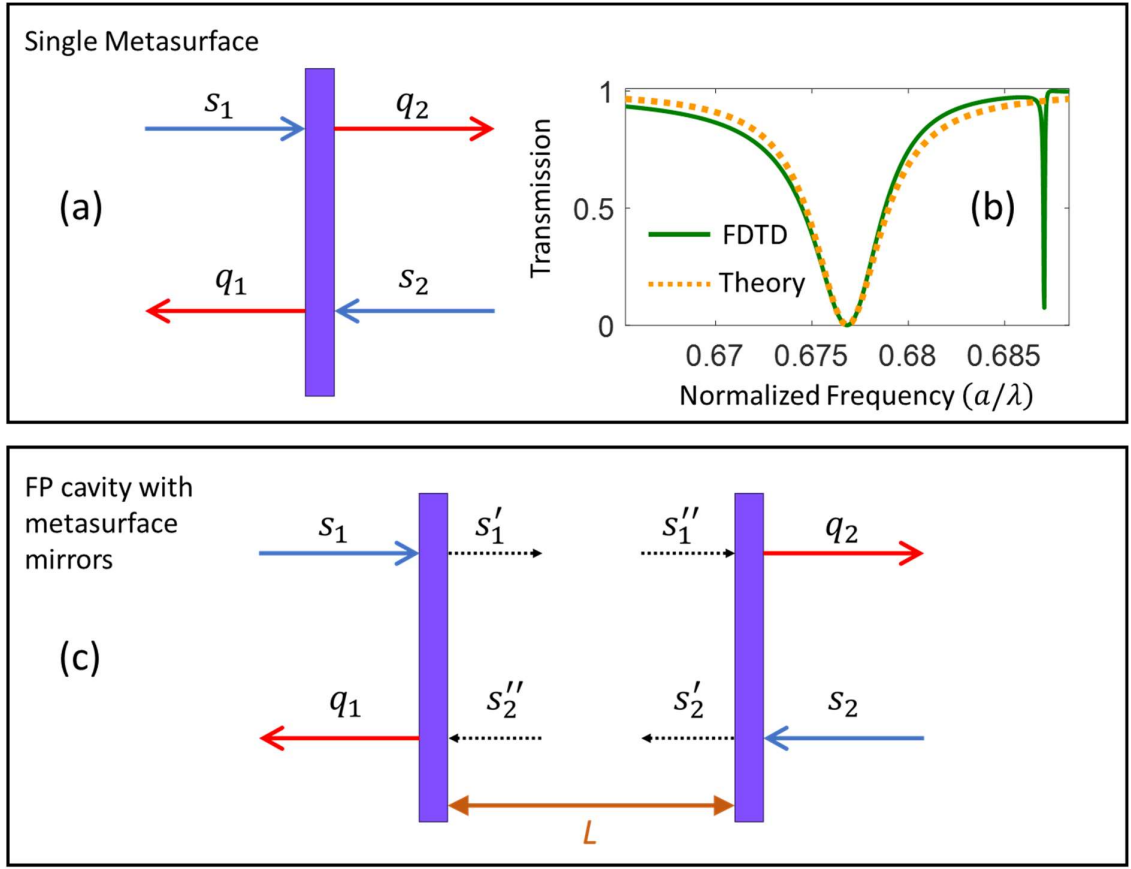

Fig. 1: (a) Schematic of a metasurface with incoming and outgoing electric field amplitudes. (b) Transmission spectrum of the metasurface with a resonance at normalized frequency $a/\lambda_s = 0.6768$. The metasurface is made of a square lattice of circular silicon nanodisks ($r$ = 150 nm, $h$ = 220 nm, $a$ = 1 μm). The ambient medium is silica. In this figure green solid line and yellow dotted line represent 3D FDTD calculation and analytical calculation (see text), respectively. (c) Schematic of a Fabry-Perot cavity made of meta-mirrors in (a).

Based on previous discussions regarding single metasurfaces, we deduce that: $s'_2 = e^{j\gamma}s_2 + c_a A_2$. Similarly, we can derive expressions for $s'_1$ and $s''_2$, By eliminating all the internal amplitiudes ($s'_1$, $s'_2$, $s''_1$ and $s''_2$), we obtain the following equations describing the dynamics of the electric field amplitude in the bilayer metasurface system:

$$\frac{dA_1}{dt} = (j\omega_s - \Gamma)A_1 + c_s s_1 + c_s\left[e^{-jkL}\left(e^{j\gamma}s_2 + c_a A_2\right)\right] + j\mu A_2 \quad [1]$$

$$\frac{dA_2}{dt} = (j\omega_s - \Gamma)A_2 + c_s s_2 + c_s\left[e^{-jkL}\left(e^{j\gamma}s_1 + c_a A_1\right)\right] + j\mu A_1 \quad [2]$$

If we assume an hypothetical system where there is no decay to the external radiation modes ($c_s = 0$), then the above equations reduce to $\frac{dA_1}{dt} = j\omega_s A_1 + j\mu A_2$ and $\frac{dA_2}{dt} = j\omega_s A_2 + j\mu A_1$. These equations can be rewritten as a second order differential equation with constant coefficients, $\frac{d^2 A_1}{dt^2} - 2j\omega_s \frac{dA_1}{dt} - (\omega_s^2 - \mu^2)A_1 = 0$, and from the solution, it can be easily shown that the resonance frequency will split into $\omega_s - \mu(L)$ and $\omega_s + \mu(L)$. Note that $\mu = \mu(L)$ is a function of $L$. For a large $L$, we have $\mu(L) \approx 0$, and therefore the splitting is not significant. Conversely, in short cavities $\mu(L)$ cannot be neglected.

In practical cases, Eqn. 1 and 2 can be solved in the Fourier domain as:

$$\begin{bmatrix} A_1 \\ A_2 \end{bmatrix} = c_s \begin{bmatrix} j\delta + \Gamma & j\mu - e^{j(\gamma-kL)}\Gamma \\ j\mu - e^{j(\gamma-kL)}\Gamma & j\delta + \Gamma \end{bmatrix}^{-1} \begin{bmatrix} 1 & e^{j(\gamma-kL)} \\ e^{j(\gamma-kL)} & 1 \end{bmatrix} \begin{bmatrix} s_1 \\ s_2 \end{bmatrix} \quad [3]$$

The outgoing electric field from the left of the bilayer metasurface is $q_1 = e^{j\gamma}s_2'' + c_a A_1 = e^{j(2\gamma-kL)}s_2 + e^{j(\gamma-kL)}c_a A_2 + c_a A_1$. Similarly, the outgoing electric field from the right of the bilayer metasurface is given by $q_2 = e^{j\gamma}s_1'' + c_a A_2 = e^{j(2\gamma-kL)}s_1 + e^{j(\gamma-kL)}c_a A_1 + c_a A_2$. If we assume the incident wave amplitude is $s_1$, then the power transmission coefficient $T = T(\omega, L)$ can be derived using $T = \left|\frac{q_2}{s_1}\right|^2\Big|_{s_2=0}$. The resulting expression can be evaluated by substituting the expressions of $A_1$ and $A_2$, and defining a normalized evanescent wave coupling coefficient $v(\omega, L) = \frac{\mu(L)}{\sqrt{\delta^2 + \Gamma^2}}$. After some long algebraic and trigonometric manipulations [see supplementary information], it can be shown that,

$$T = \frac{\{|t_s|^2 - vG(v,L)\}^2}{\{|t_s|^2 - vG(v,L)\}^2 + 4|r_s|^2\{\sin(kL - \theta_s) - v\}^2} \quad [4]$$

Here $G(v, L) = v - 2\sin(kL - \gamma)$, $\theta_s(\omega) = \arg(t_s) - \arg(\delta) + \pi/2$, and the power reflection coefficient is given by $R = 1 - T$. Note that $\theta_s$ is a smooth and continuous function of $\omega$. The discontinuity of phase in $t_s$ when $\delta = 0$ is compensated by the subtraction of $\arg(\delta)$ in the expression of $\theta_s$. Setting $v = 0$ in Eqn. 4 replicates results obtained by the scattering matrix method [26-27], which ignores evanescent wave coupling and assumes plane-wave-like propagation. For $v = \theta_s = 0$, Eqn. 4 returns to the standard transmission equation of a Fabry-Perot etalon with solid mirrors.

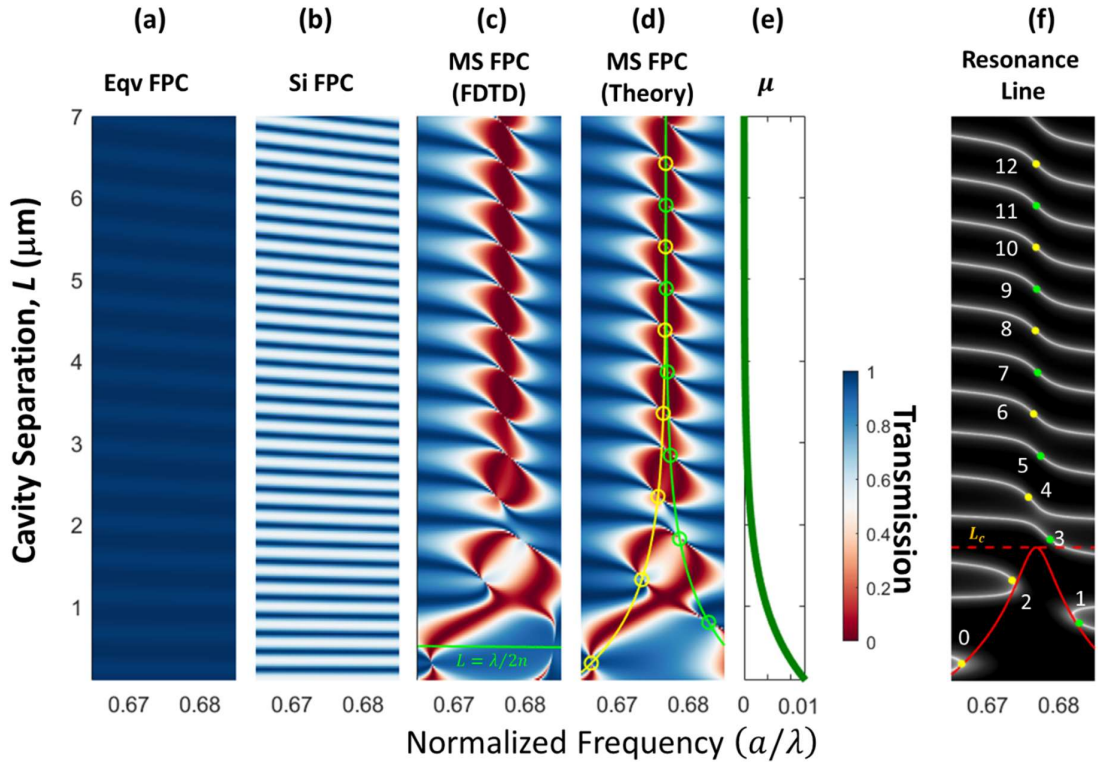

Fig. 2 Transmission spectra of Fabry-Perot cavity with various mirrors of equal thickness. (a) The mirror is a uniform dielectric layer of permittivity to volume average permittivity of the metasurface in Fig. 1(b). (b). The mirror is a silicon layer. (c) and (d) The mirror is a metasurface as in Fig. 1(b). Spectra in (c) and (d), represent 3D FDTD and analytical (Eqn. 4) calculations, respectively. (e) Coupling coefficient [$\mu(L)$] that accounts for the evanescent coupling. In (d), yellow lines represent $\omega_s \pm \mu(L)$, and the circles represent singular points. (f) White lines: Resonance lines obeying $\sin(kL - \theta_s) = v$. Dotted red line: $L = L_c$; Solid red line: locus of $\mu^2(L) = \delta^2 + \Gamma^2$. Dots with numbers: singular point, and the order $m$ [yellow: even $m$; green: odd $m$].

Let us consider FP cavities made of silicon metasurface [as described in Figs. 1(a)-1(b)] as mirrors. For comparison, Figures 2(a)-2(b), show the transmission spectra as a function of $L$ for FP cavities with a dielectric film of same thickness as metasurface as mirrors. In Fig. 2(a), the film has a dielectric constant equal to the volume-averaged dielectric constant of the metasurface, and in Fig. 2(b), the dielectric film is a silicon film, which has a higher permittivity than the volume-averaged permittivity of the metasurface. Figures 2(c)-2(d) exhibit the transmission spectra [in the vicinity of $\omega_s$] of the FP meta-mirror cavity. In Fig. 2(c), we used a 3D FDTD calculation, and in Fig. 2(d), we evaluated Eqn. 4. The evanescent wave coupling coefficient $\mu$ [Figure 2(e)] and the phase $\gamma \cong 0.6\pi$ are obtained by a numerical fitting method [28]. Clearly, both Figs. 2(c) and 2(d) share a very good agreement. There is a minor discrepancy between the FDTD and semi-analytical calculations for $\omega a/(2\pi c) > 0.68$ and $L < 0.85$ µm. This discrepancy is due to the presence of another resonance of a single metasurface in this frequency neighbourhood at $\omega a/(2\pi c) = 0.687$ [see the green line in the left panel of Fig. 1(a)] which is not captured by Eqn. 4.

As we can see from Figs. 2(a)-2(b), the transmission spectra with dielectric thin film mirrors exhibit a strips-like, periodic patterns of shallow ripples. The depth of the ripple increases with the increase in the dielectric constant of the mirror. On the other hand, the transmission spectrum of the FP cavity with meta-mirrors significantly deviates from the strip-like pattern. For longer $L$, where $v = 0$, there is a conservation of translational symmetry along the direction of $L$ variation. In this scenario, a periodic pattern is observed in the transmission spectrum. Figs. 2(c)-2(d) show that FP cavities with metasurface mirrors exhibit periodic transmission lobes instead of strips as seen in the case of dielectric film mirrors. Moreover, the variation in transmission near the lobe region is very sharp, unlike the shallow ripple-like variations seen in the equivalent FP cavities of dielectric film mirrors. For shorter $L$, we have $v > 0$, leading to the disruption of translational symmetry. Consequently, in Figs. 2(c)–2(d), we observed a non-periodic transmission pattern for shorter lengths of FP cavities with metasurface mirrors. In Fig. 2(d), we have also plotted the hypothetical lines $\omega_s - \mu(L)$ and $\omega_s + \mu(L)$ [yellow lines in Fig. 2(d)] which clearly indicate the splitting of lobes for shorter cavities.

From Eqn. 4, it is evident that when the amplitude reflection coefficient of a single metasurface is zero (i.e., $|r_s| = 0$), the power transmission of the bilayer system is always unity. This result is trivial since none of the light is reflected at either metasurface, allowing all light to pass through. Similarly, when $|r_s| = 1$, all light is expected to be reflected, and none of the light passes through the system. However, at certain separations $L$, the reflected light at each mirror constructively interferes, producing sharp

transmission peaks with unity transmission values. From Eqn. 4, it can be deduced that, this constructive interference phenomenon occurs when,

$$\sin(kL - \theta_s) = v \quad [5].$$

Eqn. 5 serves as the generalized resonant condition for a Fabry-Perot cavity with meta-mirrors. When $v = 0$, Eqn. 5 simplifies to $2nL = \frac{\lambda}{\pi}\{m\pi + \theta_s(\omega)\}$ with $m$ being any positive integer including zero. If $v = \theta_s = 0$, the solution to Eqn. 5 replicates the standard resonance condition of the usual Fabry-Perot cavity, $2nL = m\lambda$. Figure 2(f) display the resonance lines [Eqn. 5] on the $(\omega, L)$ parameter space for the FP meta-mirror cavity discussed in Figs. 2(c) – 2(d). In general, for resonance to occur we require $\sin(kL - \theta_s) = v < 1$. Hence, $\mu^2(L) < \delta^2 + \Gamma^2$. With $\delta = 0$, we can define a characteristic length, $L_c$ such that $\mu(L_c) = \Gamma = \frac{\omega_s}{2Q_s}$. For the metasurface in Fig. 1(a), $L_c$ = 1.73 μm. In Fig. 2(f), we show the boundary $\mu^2(L) = \delta^2 + \Gamma^2$ and the line $L = L_c$.

In Eqn. 4, provided $\sin(kL - \theta_s) = v$ [Eqn. 5] is satisfied, what happens when $|t_s|^2 = vG(\omega, L)$ [i.e., both numerator and denominator vanish simultaneously]? We have singular points. Mathematically, when both numerator and denominator of Eqn. 4 vanish simultaneously it can be shown that $\delta = -\mu(L)[\cos(\gamma - kL) \pm j\sin(\gamma - kL)]$. The real solution to this equation $(\omega_m, L_m)$ exists when $2nL_m = \frac{c}{\pi\omega_m}(\gamma + m\pi)$ is satisfied so that the $\sin(\gamma - kL_m)$ vanishes. The corresponding singular frequencies $\omega_m = \omega_s - \mu(L_m)$ and $\omega_m = \omega_s + \mu(L_m)$ occurs for even and odd integer values of $m$, respectively. In the neighbourhood of these singular points, we expect ultra-high-quality factors as right at the singular point there is no decay mechanism apart from the intrinsic decay. This is consistent with the earlier discussion where we arrived at the same solutions with $c_s = 0$. In Fig. 2(e), we display the singular points of the silicon FP meta-mirror cavity. As we can see from both Fig. 2(d) and 2(e), the transmission near the singular points changes rapidly. Also note that in the standard cavity, the fundamental mode has $m = 1$ and the cavity separation is $L = \frac{\lambda}{2n}$. In FP meta-mirror cavity, the singular point exists even for $m = 0$ with the corresponding cavity separation being $L_{m=0} = \frac{\lambda}{2n}\left(\frac{\gamma}{\pi}\right) < \frac{\lambda}{2n}$. The line $L = \frac{\lambda}{2n}$ is shown Fig. 2(c), as we can readily see from Fig. 2(c) [FDTD spectrum], resonance occurs for $L < \frac{\lambda}{2n}$ in the FP meta-mirror cavity.

For long cavities such that $L \gg L_c$ [i.e., $v = 0$], the equation of the resonance lines are $2nL = \frac{\lambda}{\pi}\{m\pi + \theta_s(\omega)\}$ or equivalently $L = \frac{c}{n\omega}[m\pi + \theta_s(\omega)]$. With $\omega$ being a slowly varying function compared to $\theta_s(\omega)$ the shape of $L$ versus $\omega$ curve follows very closely to the shape $\theta_s(\omega)$ [see resonance lines for $m > 5$ in Fig. 2(f)]. The resonance frequencies ($\omega_r$) for any given $L$ and order $m$ can be calculated nonlinearly from the resonance line. In the vicinity of $\omega = \omega_s$, we can linearize $\theta_s(\omega)$ as $\theta_s(\omega) = \theta_s(\omega_s) + (\omega - \omega_s)\tau_g(\omega_s)$, where $\tau_g(\omega_s) = \frac{d\theta_s(\omega)}{d\omega}\big|_{\omega=\omega_s} = -\frac{2Q_s}{\omega_s}$ is the group delay. Hence, near the $m$-th singular point, we can use the following linear relationship,

$$2nL = \left(m + \frac{\gamma}{\pi}\right)\lambda + \frac{2Q_s}{\pi}(\lambda - \lambda_s) \quad [6]$$

Right at $\lambda = \lambda_s$, we have the singular point, and Eqn. 6 can also be written as $L = L_m + \frac{Q_s}{\pi n}(\lambda - \lambda_s)$. For the transmission characteristics in the case $v = 0$, we can apply a small angle approximation, $\sin(kL - \theta_s) \approx kL - \theta_s(\omega) - m\pi$ to simplify power transmission in Eqn. 4. This approximation yields $T(\omega, L) = \frac{|t_s(\omega)|^4}{|t_s(\omega)|^4 + 4|r_s(\omega)|^2[kL - \theta_s(\omega) - m\pi]^2}$. In the vicinity of $\omega \approx \omega_r$, $|t_s(\omega)|^2 \approx |t_s(\omega_r)|^2$, and $\theta_s(\omega) = \theta_s(\omega_r) + (\omega - \omega_r)\tau_g(\omega_r)$. Thus when $v = 0$, these results allow us to rewrite $T$ in the vicinity of resonance as

$$T(\omega, L) = \frac{|t_s(\omega_r)|^4}{|t_s(\omega_r)|^4 + 4|r_s(\omega_r)|^2\left[\frac{nL}{c} - \tau_g\right]^2 [\omega - \omega_r]^2} \quad [7],$$

Eqn. 7 has a Lorentzian line shape, with the quality factor for resonant peaks given by:

$$Q = \sqrt{F}\left[\frac{\pi nL}{\lambda_r} + Q_s\right] = Q_L + Q_M \quad [8]$$

Here, $F = 4|r_s(\omega_r)|^2/|t_s(\omega_r)|^4$ is the coefficient of finesse. $F$ is a huge number that traditionally measures the sharpness of the interference fringes. In Eqn. 8, The first term $Q_L = \frac{\pi nL}{\lambda_r}\sqrt{F}$ represents the length dependant quality factor of the FP cavity. This is also the total quality factor for the standard FP cavity. The second term $Q_M = Q_s\sqrt{F}$ is unique for meta-mirrors and it is length independent. From Eqn. 8, we can also define a second characteristic length $L_Q > L_c$ that signifies the shift in quality factor domination from $Q_M$ to $Q_L$. In very long cavities such that $L \gg L_Q = \lambda_s\frac{Q_s}{\pi n}$, $Q_L$ dominates with $Q \approx Q_L$.

Conversely, in shorter cavities such that $L_c < L < L_Q$ we have $Q_M \gg Q_L$, and hence $Q_M$ dominates with $Q \approx Q_M$. For metasurface depicted in Figs. 1(a)-1(b), we have and $L_Q \cong 50$ μm.

The resonance in the case $v = 0$ of a bilayer system creates a distinct induced transparency configuration [29-30]. Utilizing bilayers of resonant metasurfaces, each exhibiting transmission dips, allows us to induce transmission peaks (as opposed to dips). Significantly, when $L_c < L < L_Q$, the quality factor of the transmission peak is proportional to $Q_s\sqrt{F}$. This is in contrast to the traditional FP cavity, where the quality factor of the resonance peak scales with the length of the cavity. Also note that as $\omega_r \to \omega_s$, $t_s \to 0$, and hence the coefficient of finesse $F$ increases sharply. Quality factor of single metasurface $Q_s$ can be engineered to be very high for single metasurface. In literature, there are plenty of reports with single metasurface of quality factors ranging between $10^2 - 10^5$. Thus, with an additional multiplication factor $\sqrt{F}$, the bilayer system can produce resonant transmission peaks with ultra-high quality.

In Figure 3(a), we illustrate the enlarged version of the resonance line for $m = 8$. As shown in this figure, in the vicinity of the singular point [closed circle in Fig. 3(a)] the linear relationship [Eqn. 6] can be used to determine the resonance condition. The transmission spectra for a discrete set of cavity separations [identified with colored, open circles Fig. 3(a)] are shown in Figure 3(b). In this figure, the induced transparency peaks are indicated with the same-coloured dots as in Fig. 3(a). Figures 3(c) and 3(d) represent transmissions as functions of $L$ for two specific resonance frequencies. These frequencies are indicated in Fig. 3(a) and 3(b) with green- and purpled-coloured dots. From Figs. 3(c) and 3(d), we can observe a periodic pattern in the resonance peaks when $L$ is varied. The length spacing between two consecutive peaks is $\frac{\lambda_r}{2n}$. In the inset of Fig. 3(b), we show the transmission spectra in the vicinity of the singular point using a finer step of $L$. As we can see from this figure, in the vicinity of the singularity, the line shapes of the transmission functions are Lorentzian. In the inset of Fig. 3(b), the Lorentzian approximations (Eqn. 7) are shown in red colors. The quality factors of these transmission peaks are depicted in Figure 3(e), where the red line and blue circles represent the analytical expression [Eqn. 8] and numerical evaluation, respectively. For the numerical calculation, we employ scattering matrix method [26], which proves to be efficient in resolving finer $L$ and corresponding frequency linewidth of the transmission peaks. The scattering matrix parameters are derived from a single metasurface using 3D FDTD calculations [27]. From Fig. 3(e), we can clearly see the good agreement between the analytical and numerical evaluations. As anticipated by the theory, the quality factor experiences a substantial increase as the resonance frequency of the bilayer system approaches the singular frequency [Fig. 3(e)]. While the quality factor of the transmission dip [Fig. 1(a)] of the single metasurface is $Q_s \cong 150$, the quality factor of the induced transparency peak in the vicinity of resonance surpasses $10^5$ [Fig. 3(e)]. Let us compare the quality factors of the meta-mirror cavities with the traditional cavities of the same coefficient of finesse. In Figure 3(f) we show quality factor of both cavities as a function of cavity separation. For the traditional cavity, the mirror has 95% power reflectivity. In the meta-mirror cavity, the frequency detuning is adjusted to give the same power reflectivity. As we can readily see from this figure, quality factors of meta-mirror cavities are significantly larger, despite a shorter cavity length. For an illustration when $a/\lambda = 0.6764$, if we use meta-mirror cavity, to achieve a quality factor $\sim 8 \times 10^3$, a cavity lengths of ~4 μm is required. However, if the same quality factor to be achieved in the traditional cavity of the same coefficient of finesse, we would require a cavity of length ~80 μm [20-fold times longer].

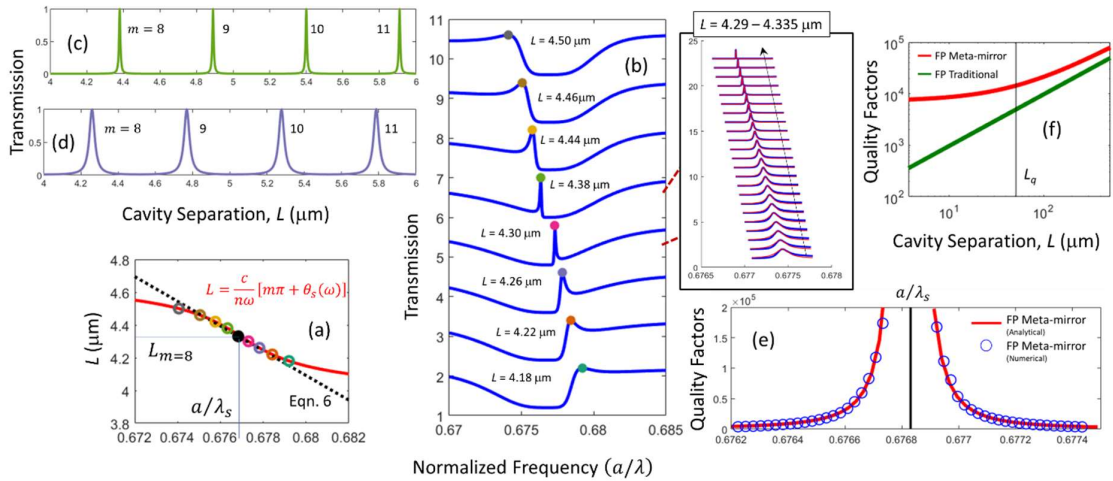

Fig. 3 Transmission characteristics of long cavities with $v = 0$. (a) Red line: resonance line for $m = 8$; Dotted line: linear approximation [Eqn. 6]; Discrete cavity separations, and the corresponding resonant frequencies $\omega_r = a/\lambda_r$ are shown in colored dots. (b) Transmission spectra for discrete cavity separations identified in colored dots as in (a); In (b), the location of resonance peaks is identified with the similar colored dots as in (a).

(c) and (d) represent transmissions as functions of $L$ for two frequencies. These frequencies are indicated in (a) and (b) with green- and purple-colored dots. (e) Quality factors as a function of resonance frequency. The corresponding transmission spectrum for a finer range of $L$ is shown in the inset of (a). (f) Comparison of quality factors for meta-mirror and traditional FP cavities as a function of cavity separation.

For shorter cavities ($L < L_c$) $v$ is significant, and thus they exhibit different resonance [Eqn. 5] and transmission [Eqn. 4] characteristics. In the first place, the singularity points deviate from $\omega_m = \omega_s$ to $\omega_m = \omega_s \pm \mu(L_m)$. For $L < L_c$, the resonance ceases to occur for $\omega = \omega_s$ as $\sin(kL - \theta_s) > 1$. Thus, the resonance line breaks and become discontinuous at $\omega = \omega_s$ [see Fig. 2(f)] giving a significantly different shape when compared to those in $L > L_c$. A resonance free region in $(\omega, L)$ parameter space is created in the neighbourhood of $\omega = \omega_s$ when $\mu^2(L) > \delta^2 + \Gamma^2$ [see Fig. 2(f)]. When $L < L_c$, the singular points and the resonance shifts to the region $\mu^2(L) < \delta^2 + \Gamma^2$. For cavities in this region, we observed sharp Fano resonances. Let $\omega = \omega_r$ to be the solution $\sin(kL - \theta_s) = v$ [Eqn. 5]. At this frequency, we have $T(\omega_r) = 1$ (i.e., a peak). If $\omega = \omega_{r'} \neq \omega_r$ is the solution $|t_s|^2 = vG(v, L)$ [i.e., vanishing numerator of Eqn. 4], then we have $T(\omega_{r'}) = 0$ (i.e., a dip). If $\omega_r - \omega_{r'} > 0$, then in the transmission spectrum a transmission peak will be followed by a transmission dip. Transmission toggles the other way around if $\omega_r - \omega_{r'} < 0$. When, $\omega_r$ and $\omega_{r'}$ come close to each other, this directional transmission switching occurs in a narrow frequency region, resulting in a Fano resonance. As the discrepancy between two frequencies shrinks, the sharpness of the resonance increases, and finally when $\omega_r = \omega_{r'} = \omega_m$, we have a singularity. In Figure 4, we plot the transmission spectra in the vicinity of singularity point with $m = 2$. As we can see from this figure, the transmission spectra of bilayer metasurfaces exhibit Fano resonances. In the same figure, we also plot the two frequencies $\omega_{r'}$ and $\omega_r$ as a function of $L$. These frequencies are nonlinearly solved from the aforementioned conditions. As we clearly observe from this figure, the sharpness of the Fano resonance increases as the two frequencies approach each other.

Figure 5(a) showcases a schematic summary of the characteristics of meta-mirror FP cavities for various cavity length regimes. For $L < L_c$ where $\mu(L) \neq 0$, mode splitting occurs. There is no resonance for $\mu^2(L) > \delta^2 + \Gamma^2$. On the other way, when $\mu^2(L) < \delta^2 + \Gamma^2$, we have Fano resonances governed by peak and valley conditions. For $L > L_c$ where

$\mu(L) \approx 0$, we have induced transparency peaks with near Lorentzian resonance lineshapes. In this regime, FP meta-mirror cavities can be further categorized according to their quality factor dominance. When $L < L_Q$, $Q_M$ dominates, and when $L > L_Q$, $Q_L$ dominates. Note that, in FP meta-mirror cavities resonances happens even for cavity separations that is smaller than the cavity length of the fundamental mode $\left[L = \frac{c}{2n\omega}\right]$ in the traditional cavity.

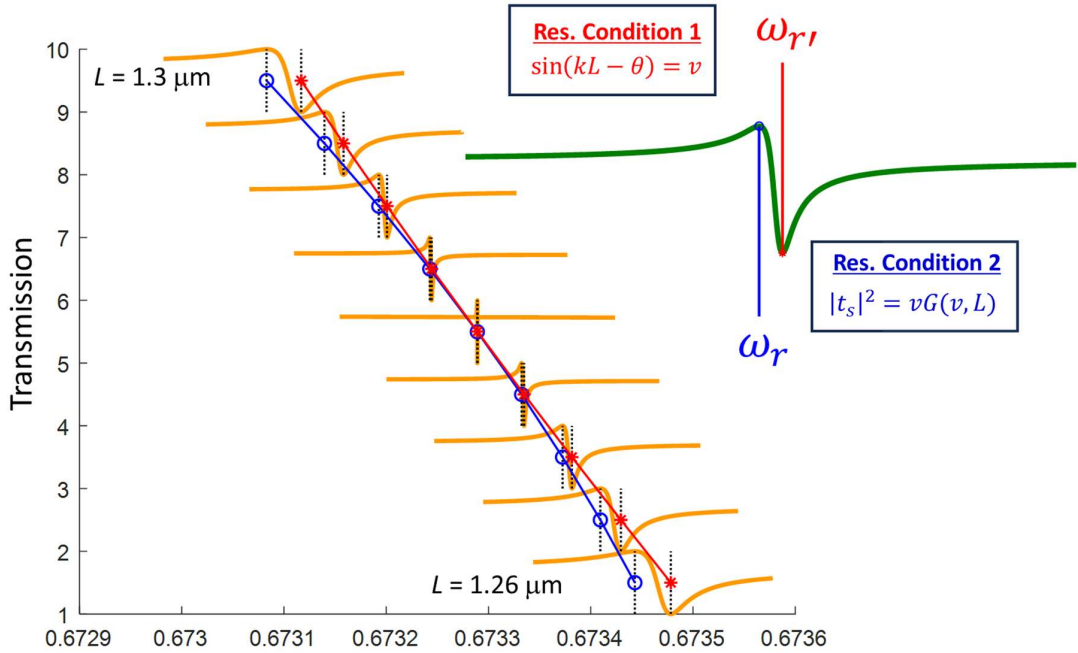

Fig. 4 Transmission characteristics of short cavities with $v > 0$. The green line schematically displays the two resonance characteristics described by Eqn. 4 (see text). The locations of $\omega_{r\prime} \approx \omega_r$ are shown with blue and red lines, respectively.

Finally, let us present the localized field distribution of the FP meta-mirror cavity. In Figure 5(b), we display the electric and magnetic field densities for both short ($L = 1.2\ \mu m < L_c$) and long ($L = 4.3\ \mu m > L_c$) cavities. For the shorter cavity [$L = 1.2\ \mu m$], the field pattern clearly shows a stronger evanescent wave interaction. Nevertheless, in both cases, the field concentrates primarily on the metasurface region rather than at the middle of the cavity. The significant group delay exhibited by the metasurface leads to a notable shift in the concentration of the electromagnetic field from the centre of the FP cavity towards the metasurface region. As demonstrated earlier in the article, this shift contributes to a remarkable enhancement in the quality factor of the FP resonance.

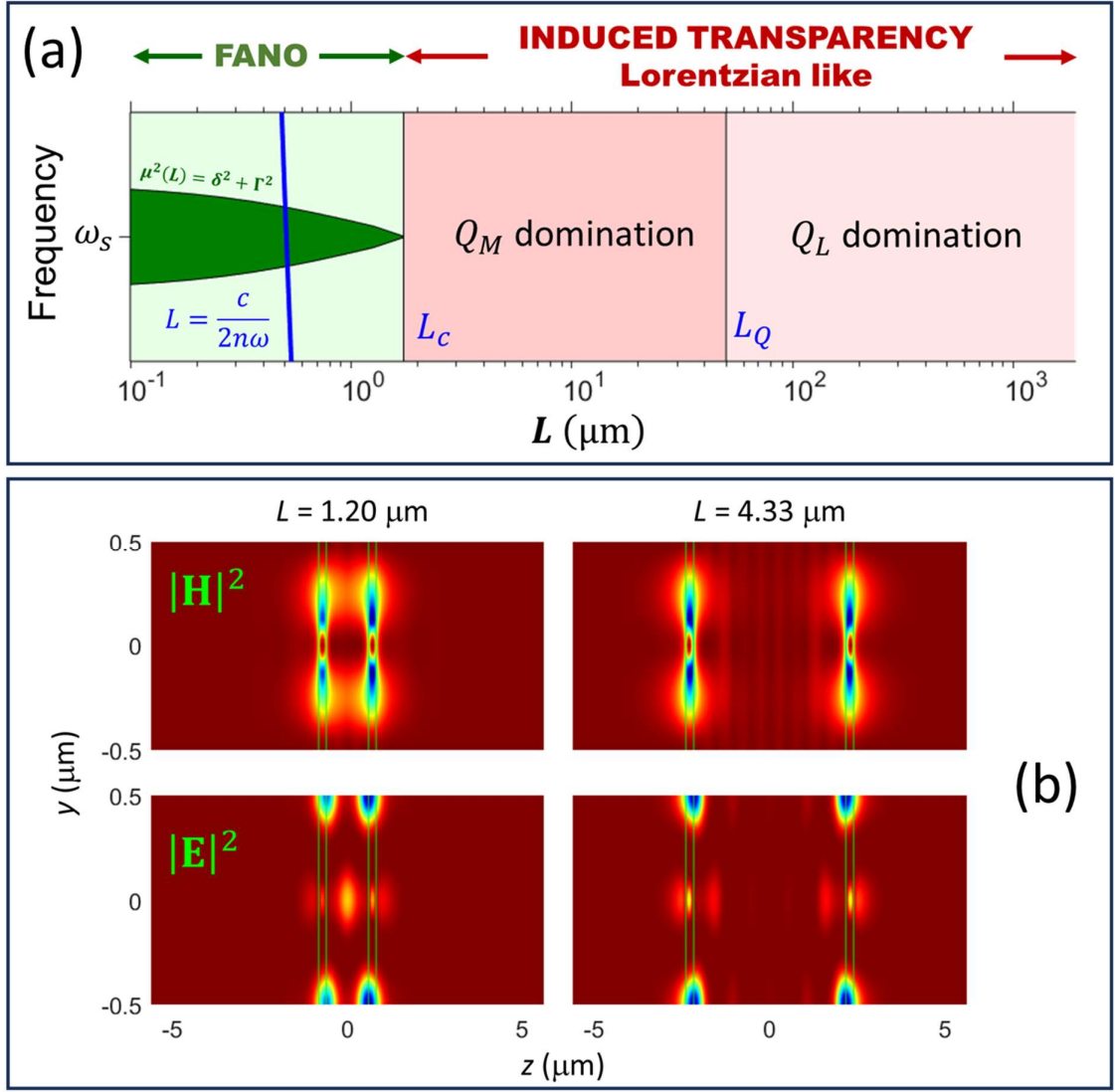

Fig. 5 (a) Schematic summary: categorization of FP cavities based on resonance line-shape, quality factors and two characteristics lengths, $L_c$ and $L_q$. The numerical values of the characteristics lengths are based on the metasurface in Fig. 1(a). (b) Electric and magnetic field densities of FP meta-mirror cavities.

In conclusion, we have presented optical properties of a system of bilayer resonant metasurfaces. This system mimics a traditional Fabry-Perot cavity with the solid mirrors replaced with meta-mirrors. We generalize the coherent interference condition for long and short cavities. When the resonant metasurface acts as meta-mirrors, the field confines to the meta-mirrors' region rather than at the middle of the Fabry-Perot cavity. We obtain an ultra-high quality factor resonance peak despite shorter cavity length, which is unattainable in the conventional Fabry-Perot cavity. We discover two characteristic cavity separations that differentiate the resonance in terms of lineshapes and dominance of quality factor. We show that the sharpness of the resonance peaks increases drastically as the resonance frequency of the cavity approaches the singular frequency.